\documentclass[prl,twocolumn,superscriptaddress,floatfix]{revtex4-1}

\usepackage{graphicx} 
\usepackage{color} 
\usepackage{amsmath} 
\usepackage[urlcolor=blue, hyperindex, colorlinks, bookmarks=true,linkcolor=black,citecolor=black]{hyperref} 
\usepackage{amssymb}
\usepackage{amstext}
\usepackage{amsthm}
\usepackage{amsfonts}

\newcommand{\dg}{^\dagger}

\newcommand{\bra}[1]{\langle{#1}|}
\newcommand{\ket}[1]{|{#1}\rangle}

\renewcommand{\section}[1]{{\em #1}.---}

\begin{document}

\title{Robust randomized benchmarking of quantum processes}
\date{\today}
\author{Easwar Magesan}
\affiliation{Institute for Quantum Computing and Department of Applied Mathematics, University of Waterloo, Waterloo, Ontario N2L 3G1, Canada}	
\author{J. M. Gambetta}
\affiliation{Institute for Quantum Computing and Department of Applied Mathematics, University of Waterloo, Waterloo, Ontario N2L 3G1, Canada}
\author{Joseph Emerson}
\affiliation{Institute for Quantum Computing and Department of Applied Mathematics, University of Waterloo, Waterloo, Ontario N2L 3G1, Canada}	
	
\begin{abstract}
In this Letter we describe a simple randomized benchmarking protocol for quantum information processors and obtain a sequence of models for the observable fidelity decay as a function of a perturbative expansion of the errors. We are able to prove that the protocol provides an efficient and reliable estimate of an average error-rate for a set operations (gates) under a general noise model that allows for both time and gate-dependent errors. We determine the conditions under which this estimate remains valid and illustrate the protocol through numerical examples.
\end{abstract}
\maketitle

The challenge of characterizing the level of coherent control over a quantum
system is a central problem in contemporary experimental physics and a fundamental task in the design of quantum information processing devices.
Full characterization of any quantum process is possible though quantum process tomography (QPT) \cite{NC97}. This has been successfully applied to the measurement of up to three coupled qubits (two-level systems) in NMR \cite{Childs2001,Weinstein04}, linear optics \cite{O'Brien2004}, atomic ions \cite{Riebe2006} and superconducting qubits \cite{CGT,Bialczak2009}. However, QPT sufferers from two shortcomings: the first is the often unrealistic assumption that the set of measurements and state preparations admit much lower errors than the process which is being characterized; the second is that the number of experiments required grows exponentially with the number of qubits, and hence QPT becomes infeasible in practice for systems consisting of more then just a few qubits.

Recently there has been significant interest in scalable methods for partial characterization of the noise affecting a quantum process~\cite{ESMR,SMKE,BPP}. In particular, randomized benchmarking (RB) protocols ~\cite{EAZ,LLEC,KLRB} have been conjectured to provide a means for characterizing the level of coherent control over a set of quantum transformations, or gates, in a way that overcomes the two shortcoming of QPT identified above. The practical feasibility of these procedures has lead to initial experimental implementations of RB in atomic ions for different types of traps \cite{KLRB,Biercuk2009}, NMR \cite{RLL}, superconducting qubits \cite{CGT,Chow2010a}, and atoms in optical lattices \cite{Olmschenk2010}.    In such protocols one simply measures the exponential decay rate of the fidelity as a function of the sequence length for random sequences of gates. The measured decay rate is presumed to give an estimate of the average error probability per gate \cite{EAZ,LLEC,KLRB,DCEL}.
However, it is easy to show (via a counter example with gate-dependent errors that consist of the exact inverse of the gate applied) that the decay rate estimated via RB methods can be totally unrelated to the actual error-rate.

In this Letter we develop a perturbative expansion for the error superoperators that leads to a sequence of increasingly precise fitting models for the experimental data, requiring only that the \emph{variation} in the errors over the RB gate set is not too strong. Our analysis is valid for a realistic noise model admitting time-dependent and gate-dependent errors and also accounts for state preparation and measurement errors.  We derive an explicit first-order fitting model for estimating the average error-rate, as well as the degree of gate-dependence in the errors associated with the RB gate set.

A RB protocol consists of the following steps:

\emph{Step 1}. Generate a sequence of $m+1$ quantum operations with the first $m$ operations chosen uniformly at random from some group $\mathcal{G} \subseteq U(d)$ and the final operation ($m+1$) chosen so that the net sequence (if realized without errors) is the identity operation. We are primarily interested in the case where $\mathcal{G}$ corresponds to the Clifford group on $n$-qubits ($d= 2^n$) because  each element of the Clifford group can be realized efficiently on a quantum processor, and because the required correction operation can be pre-computed efficiently~\cite{Got97}. In practice each operation $\mathcal{C}_{i_j}$ will have some associated error and the entire sequence can be modeled by the operation
\begin{equation}\label{eq:seq}
\mathcal{S}_\mathbf{i_m} = \bigcirc_{j=1}^{m+1}\left(\Lambda_{i_{j},j}\circ\mathcal{C}_{i_j}\right),
\end{equation}
where $\mathbf{i_m}$ is the $m$-tuple $(i_1,...,i_m)$ and $i_{m+1}$ is uniquely determined by $\mathbf{i_m}$. In the above, $\Lambda_{i_{j},j}$ is a linear superoperator representing the error (e.g., a completely positive trace-preserving map) associated with the operation $\mathcal{C}_{i_j}$, and is allowed to depend independently on the time-step $j$.  This is a very general noise model -- the only assumption is that the correlation time of the environment is negligible on time-scales longer than the time of the operation $\mathcal{C}_{i_j}$.

\emph{Step 2}. For each sequence the survival probability $\mathrm{Tr} [ E_\psi \mathcal{S}_\mathbf{i_m} (\rho_\psi)]$ is measured. Here $\rho_\psi$ is the initial state taking into account preparation errors and $E_\psi$ is the POVM element that takes into account measurement errors. In the ideal noise-free case $\rho_\psi=E_\psi=\ket{\psi}\bra{\psi}$.

\emph{Step 3}. Average over random realizations of the sequence to find the
averaged sequence fidelity,
\begin{equation}\label{eq:seqfid}
F_\mathrm{seq}(m,\psi) =  \mathrm{Tr} [ E_\psi \mathcal{S}_m  ( \rho_\psi)],
\end{equation} where
\begin{equation}\label{eq:aveS}
\mathcal{S}_m = \frac{1}{ |\{ \mathbf{i_m} \}| } \sum_{\mathbf{i_m}}^{ |\{ \mathbf{i_m} \}| } \mathcal{S}_\mathbf{i_m}
\end{equation} is the average sequence operation.

\emph{Step 4}. Fit the results for the averaged sequence fidelity [Eq.~\eqref{eq:seqfid}] to the model
\begin{equation}\label{eq:model}
F_\mathrm{seq}^{(1)}(m,\psi) =A_1 p^{m}  +B_1+C_1  (m-1)(q-p^2) p^{m-2}
\end{equation} derived below. The coefficients $A_1$, $B_1$, and $C_1$ absorb the state preparation and measurement errors as well as an \emph{edge effect} from the error on the final gate. The difference $q-p^2$ is a measure of the degree of gate-dependence in the errors, and $p$ determines the average error-rate $r$ according to the relation
\begin{equation}\label{eq:rate}
r = 1 - p - (1-p)/d.
\end{equation}
In the case of gate-independent and time-independent errors the results will fit the simpler model
\begin{equation} \label{eq:zerothorder}
F_\mathrm{seq}^{(0)}(m,\psi) = A_0 p^{m} + B_0
\end{equation}
derived below, where $A_0$ and $B_0$ absorb state preparation and measurement errors as well as an \emph{edge effect} from the error on the final gate.

The average error-rate $r$ that is determined by the above protocol has the following independent meaning.
For each error $\Lambda_{i_j,j}$ (associated with the implementation of the ideal operation $\mathcal{C}_{i_j}$), the probability of no error occurring for input state $\phi$ is just the survival probability $s_{i_j,j}(\phi) = \langle \phi | \Lambda_{i_j,j} ( | \phi \rangle \langle \phi | ) | \phi \rangle$. The (state-dependent) error probability for this operation is $ 1 - s_{i_j,j}(\psi). $ If we average this error probability over all pure input states using the invariant (Haar) measure $d\phi$, this defines a Haar-averaged error-rate for each operation
$ 1 - \overline{s_{i_j,j}} = 1 - \int d \phi \; s_{i_j,j}(\phi). $ Then, given any set of errors $\{ \Lambda_{i_j,j} \} $ associated with a set of operations $\{ \mathcal{C}_{i_j} \}$, we invoke another average over this error set in the usual way,
\begin{equation} r \equiv \frac{1}{|\{ (i_j,j) \} | } \sum_{i_j,j}^{|\{ (i_j,j) \} | } \;(1 - \overline{s_{i_j,j}}).
\end{equation}
which defines $r$, the average error-rate we want to estimate.


 The average error-rate $r$ can be determined from the observed fidelity decay in an RB experiment when the set $ \{ \mathcal{C}_{i} \}$ is a unitary 2-design \cite{DCEL} as well as a subgroup of $U(d)$. The former condition allows us to write
 \begin{equation} \overline{s_{i_j,j}}  = \frac{1}{K } \sum_l \langle \psi | \mathcal{C}_{l}^{\dagger} \circ \Lambda_{i_j,j} \circ \mathcal{C}_{l} \left( | \psi \rangle \langle \psi | \right)  | \psi \rangle, \end{equation}
where $K$ is the size of the  Clifford group on $n$-qubits. 
Direct evaluation of the average over Cliffords \cite{DCEL,DLT}, or equivalently the Haar-average \cite{Nie02,EAZ}, gives
$ \int d \psi \langle \psi | \Lambda_{i_j,j} ( | \psi \rangle \langle \psi | ) | \psi \rangle  = p_{i_j,j} + (1-p_{i_j,j})/d,$
where $p_{i_j,j}$ is a Haar-invariant (depolarization) parameter characterizing each error operator $\Lambda_{i_j,j}$.
 The average $p \equiv \frac{1}{|\{ (i_j,j) \} | } \sum_{(i_j,j)} p_{i_j,j}$ is the parameter appearing in the fitting models presented in Eqs. \eqref{eq:model} and   \eqref{eq:zerothorder}, which we now derive.


In the idealized case of gate-independent and time-independent errors we have $\Lambda_{i_j,j} = \Lambda$ for each $(i_j,j)$.  Repeated application of the identity operation $\mathcal{C}_{i_{j}}\circ \mathcal{C}_{i_{j}}\dg$ in Eq.~\eqref{eq:seq} gives
\begin{equation}
\begin{split}
  \mathcal{{S}}^{(0)}_{\mathbf{i_m}}=& \Lambda \circ \bigcirc_{j=1}^{m}\left(\mathcal{D}_{i_j}^{\dagger} \circ \Lambda \circ {\mathcal{D}_{i_{j}}}\right)
\end{split}
\end{equation} where we have used $\mathcal{C}_{i_{m+1}}\circ\cdots\circ\mathcal{C}_{i_{1}}=\openone$ and for each j defined a new gate $\mathcal{D}_{i_j}= \mathcal{C}_{i_{j}}\circ\cdots\circ \mathcal{C}_{i_{1}}$ that is independent from the gates which where performed at time-steps earlier than $j$ ($\mathcal{C}_{i_{j-1}}$ etc ). Substituting this into Eqs.~\eqref{eq:seqfid} and \eqref{eq:aveS} the average sequence fidelity is
\begin{equation}
F^{(0)}_\mathrm{seq}(m,\psi) =  \mathrm{Tr}[E_\psi \Lambda\circ \Lambda_\mathrm{twirl}^{\circ m}(\rho_{\psi})],
\end{equation} where $\Lambda_\mathrm{twirl} = \sum_{i_j}\tilde\Lambda_{i_j}/K$ with $\tilde\Lambda_{i_j}=\mathcal{D}_{i_j}\dg\circ\Lambda\circ \mathcal{D}_{i_j}$. We are left with an $m$-fold composition of gate-independent twirls over the Clifford group which implies the twirled operation $\Lambda_\mathrm{twirl} $ is  a depolarizing channel ($\Lambda_\mathrm{dep}$) \cite{DCEL}. Hence the gate-independent sequence fidelity reduces to Eq.~\eqref{eq:zerothorder} with $A_0=\mathrm{Tr}[E_\psi \Lambda(\rho_\psi-\openone/d)]$ and $B_0=\mathrm{Tr}[E_\psi \Lambda(\openone/d)]$.

More realistically the noise operator can be both gate and time-dependent $\Lambda \rightarrow \Lambda_{i_j,j}$. We can characterize the behavior of $F_\mathrm{seq}(m,\psi)$ by considering a perturbative expansion of
each $\Lambda_{i_j,j}$ about the mean error $\Lambda \equiv \frac{1}{|\{ (i_j,j) \} | } \sum_{i_j,j}^{|\{ (i_j,j) \} | }  \Lambda_{i_j,j}$. Defining
$\delta \Lambda_{i_j,j} = \Lambda_{i_j,j} - \Lambda$ $\forall i_j$, our perturbative approach will be valid provided $\delta \Lambda_{i_j,j}$ is small for each $i_j$ in a sense to be made precise later. Note that each $\delta \Lambda_{i_j,j}$ is a Hermitian-preserving, trace-annihilating linear superoperator.

Using the same change of variables described above, ie, ${\mathcal{D}_{i_m}}= \bigcirc_{j=1}^{m} \mathcal{C}_{i_j} $, we find that $\mathcal{{S}}_{\mathbf{i_m}}=\mathcal{{S}}^{(0)}_{\mathbf{i_m}}+\mathcal{{S}}^{(1)}_{\mathbf{i_m}}+\mathcal{{S}}^{(2)}_{\mathbf{i_m}}...$ where $\mathcal{{S}}^{(0)}_{\mathbf{i_m}}$ corresponds to the gate independent case, $\mathcal{{S}}^{(1)}_{\mathbf{i_m}}$ is the first order correction and so on. The first order correction consists of three terms: (1a) the small gate-dependent perturbation acts on the first gate, (1b) it acts somewhere in the middle (there are $m-1$ of these terms), and (1c) it acts on the final gate. Explicitly,
\begin{equation}
\begin{split}
\mathcal{{S}}^{(1a)}_{\mathbf{i_m}}=& \Lambda  \circ \tilde\Lambda_{i_m} \circ ...   \circ \tilde\Lambda_{i_2} \circ \left({\mathcal{D}^{\dagger}_{i_1}} \circ \delta \Lambda_{i_1,1}   \circ {\mathcal{D}_{i_1}}\right)\\
\mathcal{{S}}^{(1b)}_{\mathbf{i_m}}=&\Lambda  \circ \tilde\Lambda_{i_m}\circ ...   \circ \left({\mathcal{D}^{\dagger}_{i_j}} \circ \delta \Lambda_{i_j,j}   \circ {\mathcal{D}_{i_j}}\right) \circ ... \circ \tilde\Lambda_{i_1} \\
\mathcal{{S}}^{(1c)}_{\mathbf{i_m}}=& \delta \Lambda_{i_{m+1},m+1} \circ \tilde\Lambda_{i_m} \circ ... \circ \tilde\Lambda_{i_1}.
\end{split}
\end{equation} Averaging each of these terms over $\mathbf{i_m}$ gives
\begin{equation}
\begin{split}
\mathcal{{S}}^{(1a)}_m=& \Lambda  \circ \Lambda_\mathrm{dep}^{\circ m-1} \circ (\mathcal{Q}_1-\Lambda_\mathrm{dep}),
\end{split}
\end{equation} where we define for each $j$, $\mathcal{Q}_j=\sum_{i_j}{\mathcal{D}_{i_j}}^{\dagger} \circ \Lambda_{i_j,j}   \circ {\mathcal{D}_{i_j}}/{K}$. Note the correlations between the noise and the gate operations prevent this from resulting in a depolarized channel.

For the $m-1$ first-order terms with $ j \in \{2,...,m\}$ (case b) averaging gives $
\mathcal{{S}}^{(1b)}_m =    \sum_{j=2}^m\Lambda  \circ \Lambda_\mathrm{dep}^{\circ m-j} \circ    \left( \left(\mathcal{Q}_{j}\circ \Lambda\right)_\mathrm{dep} - \Lambda_\mathrm{dep}^{\circ2}  \right) \circ \Lambda_\mathrm{dep}^{\circ j-2}$
where the subscript ``dep" represents the depolarization of the operator within brackets. For these terms the main trick is to realize that we can re-expand ${\mathcal{D}_{i_j}} = \mathcal{C}_{i_j} \circ \mathcal{D}_{i_{j-1}} $ in order to depolarize
$ \mathcal{C}_{i_j}^{\dagger} \circ \delta \Lambda_{i_j,j} \circ\mathcal{C}_{i_j} \circ \Lambda $ under the twirling operation $ \sum_{i_{j-1}}  \mathcal{D}_{i_{j-1}}^{\dagger} \circ\cdot \circ\mathcal{D}_{i_{j-1}}/K$. Using the fact that depolarizing channels commute, $\mathcal{{S}}^{(1b)}_m$ simplifies to
\begin{equation}
\mathcal{{S}}^{(1b)}_m =  \sum_{j=2}^m\Lambda  \circ
\left( \left(\mathcal{Q}_{j} \circ \Lambda\right)_\mathrm{dep} - \Lambda_\mathrm{dep}^{\circ2}  \right) \circ  \Lambda_\mathrm{dep}^{\circ m-2}.
\end{equation}

To find the expression for $\mathcal{{S}}^{(1c)}_m$ we have to use the fact that the Cliffords are a group. If $i_1,...,i_{m-1}$ are fixed, averaging over the $i_m$ index runs through every Clifford element with equal frequency in the $\mathcal{D}_{i_{m}}$ random variable. Since $\Lambda_{i_{m+1},m+1}$ is just the error associated with the gate $\mathcal{D}_{i_{m}}^{\dagger}$, $\sum_{i_m} \delta\Lambda_{i_{m+1},m+1} \circ \left(\mathcal{D}_{i_m}^{\dagger} \circ \Lambda \circ {\mathcal{D}_{i_m}} \right)/K$ is independent of the $i_1,...,i_{m-1}$ indices and
\begin{equation}
\mathcal{{S}}^{(1c)}_m =    \left(\mathcal{R}_{m+1}-\Lambda\circ \Lambda_\mathrm{dep}\right)\circ \Lambda_\mathrm{dep}^{\circ m-1}
\end{equation} where $\mathcal{R}_{m+1}=\sum_{i_m} \Lambda_{i_m^{\prime},m+1} \circ \left(\mathcal{C}_{i_m}^{\dagger} \circ \Lambda \circ {\mathcal{C}_{{i_m}}} \right)/K$. In the above sum  $\Lambda_{i_m^{\prime}, m+1}$ denotes the error that arises when the Clifford operation  $\mathcal{C}_{i_m}^{\dagger}$ is applied at final time-step $m+1$. 

Combining these three terms it can be shown that the average sequence fidelity is given by Eq. \eqref{eq:model} with
\begin{eqnarray}
A_1&=&\mathrm{Tr}\left[E_\psi \Lambda\left(\frac{\mathcal{Q}_1(\rho_\psi)}{p}-\rho_\psi+ \frac{(p-1)\openone}{pd}\right)\right]\nonumber\\
&&+\mathrm{Tr}\left[E_\psi \mathcal{R}_{m+1}\left(\frac{\rho_\psi}{p}-\frac{\openone}{pd}\right)\right]\\
B_1&=&\mathrm{Tr}\left[E_\psi \mathcal{R}_{m+1}\left(\frac{{ \openone}}{d}\right)\right]\\
C_1&=&\mathrm{Tr}\left[E_\psi \Lambda\left(\rho_\psi-\frac{\openone}{d}\right)\right]
\end{eqnarray} where $q=\sum_{j=2}^m q_j/(m-1)$ and $q_j$ is depolarizing parameter defined by $(\mathcal{Q}_{j} \circ \Lambda)_\mathrm{dep}(\rho) = q_j \rho + (1 - q_j) \openone/d$. Here we see that the edge effects are represented by the three coefficients $A_1$, $B_1$, and $C_1$ and there is a slight $m$ dependence in the $A_1$, and $B_1$ coefficients due to the last gate. This is unavoidable but as long as $m$ is large enough the exponential dependence will be distinguished from this dependence. Furthermore, if the errors don't change as a function of time and only depend on the gate implemented (not time) then this dependence also disappears.

We now give conditions for when it is justified in stopping the expansion at first order. We use the ``$1 \rightarrow 1$'' norm on linear superoperators maximized over Hermitian  inputs, denoted $||\cdot||^H_{1\rightarrow 1}$, to make this precise~\cite{Wat05}. Applying the triangle inequality multiple times we find that for each order $k$
\begin{equation}\label{eq:gamma}
\Big{\|}S^{(k)}_m\Big{\|}_{1\rightarrow 1}^H \leq \sum_{j_k>...>j_1}\gamma_{j_k}...\gamma_{j_1}
\end{equation} where
$\gamma_{j} := \sum_i \left\|\Lambda_{i_j} - \Lambda\right\|_{1\rightarrow 1}^H/K$ is a measure of the variation in noise. In the case where the noise is time-independent Eq. \eqref{eq:gamma} becomes ${\|}S^{(k)}_m{\|}_{1\rightarrow 1}^H\leq {m+1 \choose {k}}\gamma^k$. Note that this norm bounds the fidelity and thus we have $|F^{(k+1)} - F^{(k)} | \leq \Big{\|}S^{(k)}_m\Big{\|}_{1\rightarrow 1}^H \leq {m+1 \choose {k}}\gamma^k$. Hence the $k+1$ order correction to the fidelity formula can be neglected provided that $$(m+1-k)\gamma/(1+k) \ll 1. $$ Therefore we can ignore second order terms when the variation in error strengths satisfies $\gamma \ll 2/m$. Note that in practice one also needs $m \gg 1$  in order to generate enough data points for a reasonable estimate of $p$ in the fitting model.

As an example of the procedure, we first consider the case of benchmarking a single qubit under time-independent unitary errors with no state-preparation or measurement errors. For each $\mathcal{C}_j$, the unitary error was constructed by finding the Hamiltonian that generates the Clifford operation via $\mathcal{C}_j(\rho)=\exp(-iH_j) \rho \exp(iH_j)$. For each $H_j$, the unitary $\exp(-iH_j)$ was diagonalized and to simulate the error one of the eigenvalues was multiplied by $e^{i\delta}$ and the other by $e^{-i\delta}$. Physically this corresponds to over/under rotations around $H_j$.  

\begin{figure}\begin{center}
\includegraphics[width=.40\textwidth]{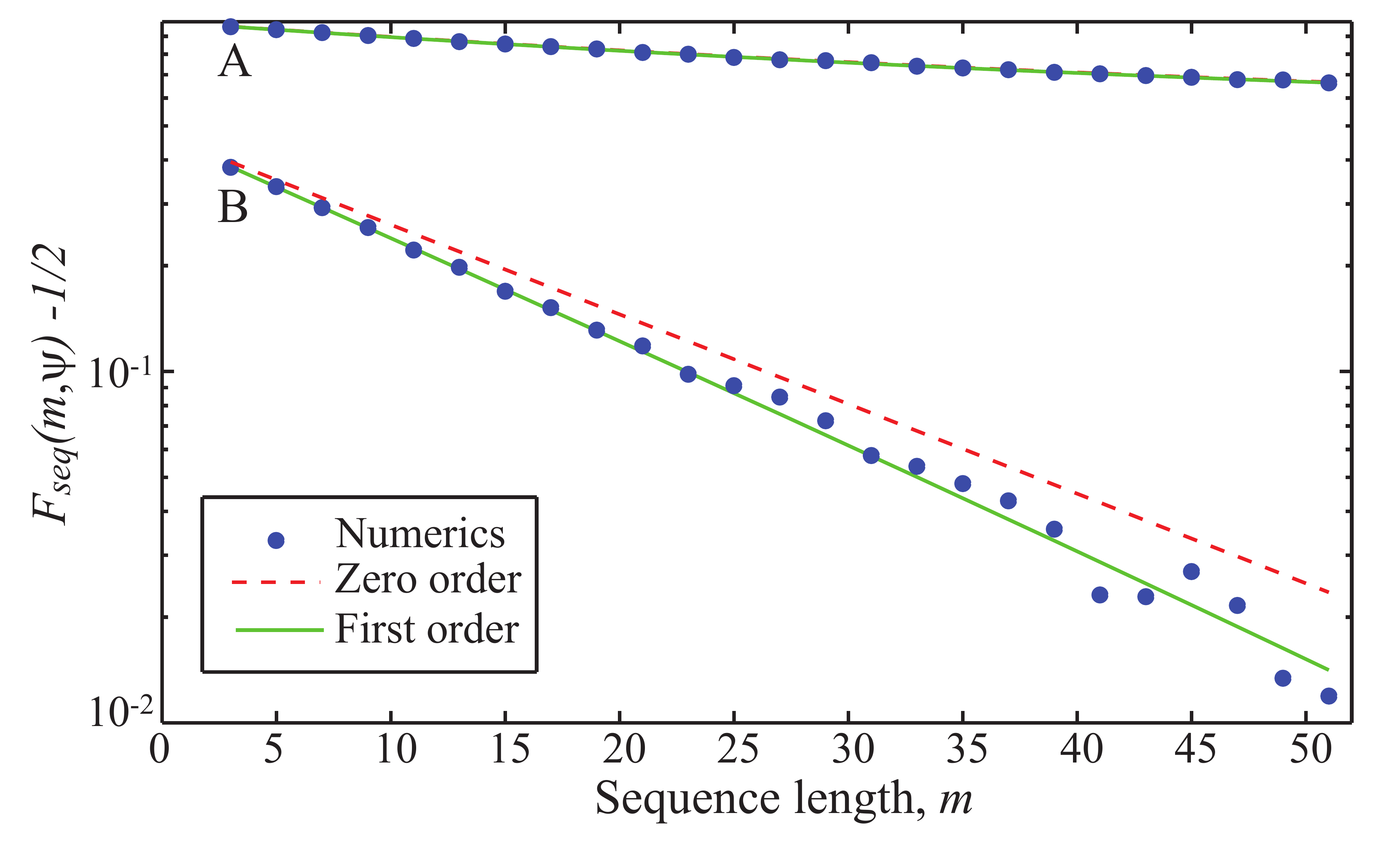}
\caption{\label{Fig:Fid}(color online) Average sequence fidelity as a function of sequence length for a error model with unitary noise. See text for details.}\end{center}
\end{figure}

Two cases for $\delta$ were analyzed: $\delta = 0.1$ (case A) and $\delta$ chosen uniformly at random in the range $[0.075,1.125]$ (case B). Numerical values for $F_\mathrm{seq}(m,\psi)$ are shown in Fig. \ref{Fig:Fid} as blue points. Note that we have subtracted the DC offset in the model so that pure exponentials appear as straight lines on the semilog plot. In the present case $B_1 = B_0 = 1/2$ since the noise is unital and there are no state preparation or measurement errors. For both cases the first order result fits the data extremely well (green line) while the zero'th order (red dashed) only approximates the sequence fidelity when the variation in $\delta$ is small (case A). Furthermore, in case B the non-exponential behaviour of the average sequence fidelity is clearly visible.   This is also apparent in Table \ref{tab:num} which shows that the gate-dependence fit parameter $q-p^2$ is much larger for case B than case A.

We also considered two other error models of practical relevance: unitary error with depolarizing noise and unitary error with amplitude damping. The depolarizing and damping parameters were chosen randomly in $0.9875 \pm 0.01$ with the unitary error chosen in the same way as case A. The results are summarized in Table \ref{tab:num} - in both these cases the simulations are well approximated by the zero'th order solution. These results further illustrate that the zero'th-order randomized benchmarking model gives a robust estimate of the error-rate for a variety of error models provided that the variation in the noise is small enough.

\begin{table}
\centering
	\begin{tabular}{l|cccc}
	\hline
\hline
   &\scriptsize{Unitary A}  & \scriptsize{Unitary B}& \scriptsize{Unitary and Dep.} & \scriptsize{Unitary and $T_1$} \\ \hline
\scriptsize{$p$} & \scriptsize{0.980} & \scriptsize{0.943} & \scriptsize{0.982} & \scriptsize{0.988} \\
\scriptsize{$r$} & \scriptsize{1.05e-2} & \scriptsize{2.85e-2} & \scriptsize{8.75e-3} & \scriptsize{5.85e-3} \\
\scriptsize{$q-p^2$}  & \scriptsize{-2.73e-4} & \scriptsize{-6.83e-3}& \scriptsize{-2.77e-8} & \scriptsize{-2.80e-8} \\ \hline\hline
\end{tabular}
\caption{\label{tab:num}Numerical results for the parameter $p$, error rate $r$, and our gate dependence measure $q-p^2$ for the four cases of noise models considered.  See text for details.}
\end{table}

The size of the Clifford group scales as $2^{O\left(n^2\right)}$~\cite{CJL} and so the number of sequences of length $m$ scales as $2^{mO\left(n^2\right)}$. Hence full averaging over the Clifford group is not efficient. Fortunately, random sampling provides an efficient means of estimating the full average. Intuitively this can be understood by realizing that the end goal is simply to estimate a single probability for each $m$, and the number of repetitions required for this should be no more than what is required to estimate the bias of a coin.  More precisely, Hoeffding's inequality specifies that, with confidence $\delta$ and accuracy $\epsilon$, the number of trials $k$ needed for approximating the average sequence fidelity is no larger then $k=\ln(2/\delta)/2\epsilon^2$, which is independent of $m$ and $n$. This result assumes one can sample uniformly from the Clifford group. For experiments on just a few qubits, one can sample uniformly from the Clifford group by constructing an exhaustive list of distinct Clifford elements and then sampling uniformly from that list. 

 For arbitrary $n$, the efficiency of our protocol relies on the ability to sample uniformly either from the full Clifford group or some approximate 2-design that is also a group, in an efficient way.  
 One method for producing approximately uniformly random elements of the full Clifford group (and hence creating an approximate 2-design), proposed in Ref.~\cite{DLT}, consists of randomly applying gates from a specific generating set G for the Clifford group. In this approach the number of gates $b$ that are required is shown to scale polynomially in $n$ when the generating set consists of C-NOT's on all pairs of qubits, and all single qubit Hadamards and phase gates (with inverses).
  
  Another approach is to take the generating set to be the full two-qubit Clifford group on each pair of qubits, in which case the number of gates needed to generate an approximate 2-design on $n$ qubits is a polynomial of substantially smaller degree \cite{HL}.  In any of these random circuit approaches the RB protocol yields an  error-rate $r$ that is associated with the cumulative error operator for a sequential block of one and two-qubit generating gates of size $b$. 
The total number of gates in the RB protocol is $m b$, where $m$ is the number of operations sampled from the corresponding approximate 2-design.
 

In conclusion we have described a protocol for estimation of error-rates in noisy quantum information processors that consists of applying random sequences of Clifford operations and measuring the average sequence fidelity. We prove that, provided the variation in the errors is not too strong, this protocol gives an efficient and reliable estimate of the average error-rate for a realistic noise model which admits both gate and time-dependent errors. We derive zero'th-order and first-order fitting models for the experimental data and numerically illustrate the relevance of both models. 

\begin{acknowledgments}
We thank Marcus Silva and David Cory for valuable discussions. E.M. was supported by NSERC and CIFAR, J.M.G. was supported by CIFAR, Industry Canada, MITACS, MRI and NSERC and J.E. was supported by NSERC, CIFAR, and an ERA grant from the Ontario government.
\end{acknowledgments}


\end{document}